\documentclass[a4paper]{revtex4}
\usepackage{graphicx}
\usepackage{fancyhdr}
\usepackage{amsfonts,amsthm}
\usepackage{amsmath,amssymb}
\usepackage{color}

\begin{document}

\title{Towards systematic exploration of multi-Higgs-doublet models}


\author{Igor Ivanov}
\affiliation{CFTP, Instituto Superior T\'ecnico, University of Lisbon, Portugal}

\begin{abstract}
Conservative bSM models with rich scalar sector, such as multi-Higgs-doublet models,
can easily accommodate the SM-like properties of the 125 GeV scalar observed at the LHC.
Possessing a variety of bSM signals, they are worth investigating in fuller detail.
Systematic study of these models is hampered by the highly multi-dimensional parameter space
and by mathematical challenges.
I outline some directions along which multi-Higgs-doublet models in the vicinity of a large
discrete symmetry can be systematically explored.
\end{abstract}

\maketitle

\thispagestyle{fancy}


%%%%%%%%%%%%%%%%%%%%%%%%%%%%%%%%%%
\section{The quest for systematic exploration of NHDM}

The LHC Run 1 results, and especially the measured Higgs boson properties \cite{LHC-run1}, 
reveal a SM-like picture of the particle physics at the electroweak scale. 
Nevertheless this situation is compatible with a potentially very rich scalar sector,
hidden from direct inspection due to some decoupling or alignment arguments.
Assuming that this is the situation at the TeV scale, one can ask
what are the best experimental shortcuts to this hidden sector, and 
how one should probe it at colliders.
Of course, the answer strongly depends on the Higgs sector chosen,
and hundreds of papers have investigated it in various specific circumstances.
Also, the full parameter space is usually huge which renders it impossible
to systematically explore the model in its full complexity, neither with algebra nor with numerical methods.
In these cases, one usually resorts to specific models in particular corners of the parameter space.

Without any direct clue from experiment, one usually tries to uncover all physically interesting situations
within a given model and to check which of them offer the most attractive description of the data.
This undertaking requires a systematic exploration of the entire parameter space of a model,
which for most cases is beyond the reach of traditional methods.

One of the main reasons is that, with multiple Higgs fields, the scalar potential, 
even the renormalizable one, becomes very complicated.
Even an efficient description of the full parameter space is a challenge,
let alone its investigation.
The two-Higgs-doublet model (2HDM) \cite{2HDM} is a hallmark example of the bSM Higgs sector,
whose phenomenology exhibits a variety of bSM effects and which, at the same time,
is still amenable to systematic treatment in the entire parameter space,
although with somewhat non-standard mathematical methods.
For more involved Higgs sectors, the complexity of the analysis skyrockets, making the systematic study impossible.
As a result, extensions of the Higgs sector beyond 2HDM, such as $N$-Higgs-doublet models (NHDM),
are nowhere near in the detail of their investigation,
despite hundreds of publications 
(for a few early publications, see \cite{3HDM-early}).
It all makes systematic investigation of NHDM a task which is very challenging but worth pursuing.

How should one attack this problem?
Experience in conservative bSM model-building shows that models, 
which are phenomenolgically viable and theoretical attractive, 
often arise from additional symmetries, either exact or approximate. 
NHDMs can have much richer (discrete) symmetries than 2HDM but usually 
these exact symmetries lead to either unrealistic or very SM-like phenomenology.
It is therefore natural to systematically explore NHDMs {\em in the vicinity of large discrete symmetry group}.

This task splits into two parts, each of them being challenging on its own. 
First, one needs to know which highly symmetric NHDMs one can construct
for a given $N$ and what are their phenomenological consequences.
If we assume that Higgs doublets transform under irredicible representation of a large symmetry group $G$,
then the renormalizable Higgs potential takes the form
\begin{equation}
V = - m^2 \sum_{i=1}^N (\phi^\dagger_i \phi_i) + V_4(G)\,,
\end{equation}
with the quadratic part being symmetric under all linear unitary transformations of $N$ doublets
and the quartic part $V_4(G)$ encoding the desired symmetry.

Second, one should explicitly break this symmetry group in a way that does not lead to dramatic
consequences in order not to violate existing data. 
A natural procedure is to introduce soft breaking terms in the quadratic
potential, $m_{ij}^2 (\phi^\dagger_i \phi_j)$, keeping the quartic part unchanged.
In this way, the symmetric model serves as a starting point, and one needs to 
systematically trace the evolution of phenomenology as one shifts away from the symmetry.

The model still remains highly multidimensional, and the direct scans of the entire parameter space
is of little use. I propose instead to search for analytically calculable {\em robust quantities}
which would be applicable to a wide class of models and
not be sensitive to specific numerical values of free parameters. 
It would be especially encouraging if these quantities are basis-invariant.

Below I will outline some directions in 3HDM, along which this task can be pursued.

\section{NHDM in the vicinity of high discrete symmetry}

\subsection{Symmetries in the 3HDM scalar sector and their breaking}

The first step is to understand the symmetric situations possible
within a given scalar sector.
Here, we use for illustration the three-Higgs-doublet model (3HDM), whose scalar symmetries
have been recently investigated in much detail in \cite{abelian,3HDMsymmetries,geometric,3HDMbreaking}.
The effect of these symmetries and their breaking on the fermionic sector were investigated
in \cite{silvaA4S4} for specific groups and, in general terms, in \cite{nogo}
which completes the old analysis \cite{Nir}.

The mere fact that we have at our disposal only three doublets, which interact via a renormalizable potential,
restricts the list of symmetry groups $G$ which can be implemented in such scalar sectors.
Limiting ourselves only to discrete groups, one obtains the following list \cite{abelian,3HDMsymmetries}:
\begin{equation}
G \quad = \quad Z_2,\quad Z_3,\quad Z_4,\quad Z_2\times Z_2,\quad S_3, \quad 
D_4,\quad A_4,\quad S_4,\quad \Delta(54)/Z_3, \quad \Sigma(36)\,.
\end{equation}
Imposing any other discrete symmetry group on the 3HDM scalar sector will unavoidably lead to
an accidental continuous symmetry. Some of these groups, namely, $Z_4$,
$D_4$, $A_4$, $S_4$, and $\Sigma(36)$, automatically lead to explicit $CP$-conservation in the 
scalar sector; the others are compatible with explicit $CP$-violation.

\begin{table}[htb]
\centering
  \begin{tabular}{| r c | c c | c |}
\hline
group & $|G|$ & $|G_v|_{min}$ & $|G_v|_{max}$ & sCPv possible? \\   \hline\hline
abelian & $2,\, 3,\, 4,\ 8$  & 1 & $|G|$ & yes \\
$ Z_3 \rtimes  Z_2^*$ & 6 & 1 & 6  & yes \\
$S_3$ & 6 & 1 & 6  & ---\\
\hline
$ Z_4 \rtimes  Z_2^*$ & 8 & 2 & 8  & no\\
$S_3\times Z_2^*$ & 12 & 2 & 12  & yes\\
$D_4\times Z_2^*$ & 16 & 2 & 16  & no\\
\hline
$A_4\rtimes Z_2^*$ & 24 & 4 & 8  & no\\
$S_4\times  Z_2^*$ & 48 & 6 & 16  & no\\
$CP$-violating $\Delta(54)$ & 18 & 6 & 6  & ---\\
$CP$-conserving $\Delta(54)$ & 36 & 6 & 12 & yes \\
$\Sigma(36)$ & 72 & 12 & 12 & no\\
\hline
\end{tabular}
\caption{The amount of residual symmetry possible after EWSB for each discrete symmetry group of the 3HDM scalar potential 
(see text for details). $ Z_2^*$ signals the presence of a (generalized) $CP$ symmetry in the model.}
\label{table}\end{table}

All possible symmetry breaking patterns for each of these groups were listed in \cite{3HDMbreaking};
see also \cite{geometric,trautner} for results in specific groups.
These findings are summarized in Table~\ref{table}. 
The strongest symmetry breaking of a given group $G$, in which we also include (generalized) $CP$-symmetries
when they are present, corresponds to the smallest residual symmetry group  $G_v$, and its order is denoted by $|G_v|_{min}$.
The weakest breaking corresponds to the largest residual symmetry group, with order $|G_v|_{max}$.
The groups in the upper block allow for all types of symmetry breaking:
complete, partial, or no breaking at all.
The groups in the middle block can remain intact at the global minimum, but if they are broken,
their breaking is only partial. 
The last block contains groups which can neither remain unbroken nor break completely.
They are always broken to a proper subgroup.

The fact the large discrete symmetry groups cannot be broken completely
with tree-level potentials has several phenomenological consequences.
In the last column of Table~\ref{table} we indicate 
whether the $CP$-symmetry present in the scalar sector can spontaneously break.
One sees that highly symmetric models prevent not only explicit but also spontaneous $CP$-violation
in the scalar sector. 
It is curious to note that these two types of $CP$-violations always come in pairs, at least in the 3HDM.
Whether this is just a coincidence or reveals a generic fact in NHDM is not yet known.

Incomplete symmetry breaking has also consequences for the fermion sector.
Working within pure NHDM (no bSM fields beyond several Higgs doublets)
and extending the symmetry group $G$ to the full theory,
one can prove that the lack of complete breaking leads to unphysical quark sector.
The presence of a residual symmetry among Higgses coupled to quarks
leads either to massless quarks, or to block-diagonal form of the CKM matrix,
or to the absence of $CP$-violation. This detrimental role of residual family symmetries
were noted long ago \cite{Nir} but the accurate statement and its proof were presented 
only recently \cite{nogo}.

NHDM with large symmetry groups contain many fields but very few free parameters,
which leads to degeneracies among some of the physical Higgses.
Examples of this situation appeared in many papers, and, for the charged Higgs bosons,
could be traced from general NHDM analysis \cite{celso}.
Residual symmetries can also stabilize some of the Higgs bosons
making them dark matter candidates. This is a well-known feature of multi-scalar models,
the Inert doublet model \cite{IDM} and the simple extensions with EW singlets \cite{singlet}
being the most studied examples. Here we want to stress that, for large discrete groups, 
such a symmetry-protected stabilization can be automatic and does not rely 
on one's choice of free parameters.

In short, NHDMs equipped with large discrete symmetry groups
have predictive phenomenological consequences, which are not sensitive
to the exact numerical values of the free parameters. 
These consequences reflect robust structural properties
of these models and especially of the scalar potential.
It is true that some of NHDMs with large discrete symmetry group 
are already incompatible with experiment, especially
in the fermon sector. However if one wants to softly break these symmetries later,
it is not an obstacle but rather a good starting point in understanding phenomenology.

\subsection{Critical exponents}

The second step is to softly break the large symmetry group and to track down 
how the phenomenological picture changes.
Certainly, for concrete models, one can perform this analysis numerically.
However understanding in this way the entire spectrum of physical possibilities
is a hopeless task due to the huge number of free parameters in multi-Higgs sectors.
If we want to gain some insight into these models in the entire parameter space,
we need to devise a method beyond simple numerical scans or case-by-case investigation.

A promising direction is to find robust and calculable quantities, 
which would not be too sensitive to the exact numerical values of the free parameters 
but which would reflect the structural features of the whole class of models.
Here I describe one example of such quantities which I call ``critical exponents''.

\begin{figure}
\includegraphics{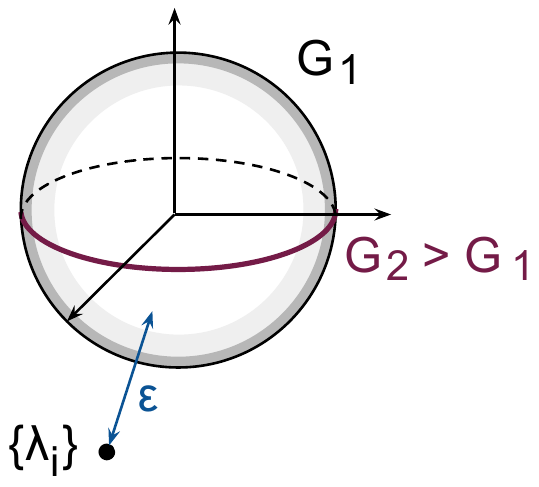} 
\caption{\label{fig-space} In the multi-dimensional parameter space, symmetric models occupy lower-dimensional
manifolds. These manifolds are embedded into one another dependeing on their symmetrry groups
$G$. The phenomenology of a non-symmetric model in the vicinity of a symmetric situation 
can bear resemblance to the symmetric case, up to corrections which are powerlike in the distance $\epsilon$.}
\end{figure}

Consider a model with large multi-dimensional parameter space $\{\lambda_i\} \subset R^n$,
see Fig.~\ref{fig-space}.
Models possessing (large) symmetry groups can be contructed by imposing 
relations among these free parameters; thus, they occupy certain low-dimensional manifolds in this space.
Their phenomenology in the scalar and fermionic sector often features quantities which are zero. 
Depending on the specific construction,
they can reflect degenerate extra Higgs bosons, their stability due to residual symmetry after EWSB,
massless quarks, absence of the $CP$-violation in the quark mixing matrix, etc.
Symmetry-based NHDM examples with these properties can be easily constructed.
We call these symmetry-protected quantities ``order parameters'' and denote them generically as $x$. 

Now, a generic point $\{\lambda_i\}$ in the parameter space does not correspond to any exact symmetry,
hence it leads to non-zero order parameters $x\not = 0$.
However this point can lie close to a symmetric manifold, within small distance $\epsilon$,
where ``small'' means comparable or smaller than typical sizes of the symmetric manifold structures.
One can then expect the order parameters to exhibit
a powerlike behaviour: $x \propto \epsilon^{\nu}$.
The index $\nu$ is called the critical exponent for the quantity $x$.
A set of critical exponents describes how the phenomenology evolves in the vicinity of a symmetric situation.

There are two important aspects of these critical exponents.
First, they are robust upon variation of free parameters and are, therefore, calculable.
They do not depend on the exact position of the model in the parameter space, but only reflect
its degrees of proximity to a certain symmetric situation.
They can depend, though, on the symmetry group and on the geometric shape of its manifold in the parameter space.
In short, critical exponents reveal certain structural properties of the whole family of models,
and should be calculable analytically.

Second, having these critical exponents at hand will provide qualitative understanding of multi-parametric models 
in the vicinity of symmetries, and in particular of NHDM with large softly broken non-abelian groups.
Indeed, knowing whether certain critical exponent is $\nu_1 = 3$ and another is $\nu_2 = 1/3$ would 
immediately provide insight on which effects are phenomenologically important and how they can be related.
In short, when working in hugely multidimensional parameter spaces,
critical exponents can guide the search for models with desired phenomenology.

As an illustration of critical exponents displaying non-trivial behaviour, consider the scalar sector in the general 2HDM 
in the vicinity of critical points in the parameter space at which the lightest Higgs mass $m_h=0$.
The parameter space can be fully reconstructed \cite{ivanovActa}, and locus of critical points corresponds to a certain ellipse in it. 
In its vicinity, the mass is non-zero and the critical exponent $\nu$ was found to be either $1/2$ or $1/3$, 
depending on how one approaches the ellipse \cite{ivanovPRE}. These two cases differ by a residual discrete symmetry of the model.
This example, not pretending to be phenomenologically relevant, provides a taste of how easily the things can become non-trivial
even in rather simple situations. 
Systematric derivation of critical exponents for all order parameters in NHDM will already provide a new insight into
its phenomenology.

\begin{acknowledgments}
This work was supported by the Portuguese fund FCT under the FCT Investigator Development grant.
\end{acknowledgments}

\bigskip % extra skip inserted

\end{document}